\documentclass{emulateapj}
\usepackage{apjfonts}
\usepackage{lscape}

\slugcomment{Accepted by the Astronomical Journal}

\shorttitle{New high proper motion stars south of Decl.=-30$^{\circ}$.}
\shortauthors{Sebastien Lepine}

\begin{document}

\title{New High Proper Motion Stars from the Digitized Sky Survey.
III. Stars with proper motions $0.45<\mu<2.0\arcsec yr^{-1}$ south of
Decl.=-30$^{\circ}$.\altaffilmark{1}}

\author{S\'ebastien L\'epine}

\affil{Department of Astrophysics, Division of Physical Sciences,
American Museum of Natural History, Central Park West at 79th Street,
New York, NY 10024, USA}

\altaffiltext{1}{Based on data mining of the Digitized Sky Survey,
developed and operated by the Catalogs and Surveys Branch of the Space
Telescope Science Institute, Baltimore, USA.}

\begin{abstract}
We report the discovery of 182 new southern stars with proper motion
larger than $0.45\arcsec$ yr$^{-1}$. The stars were found in an
expansion of the SUPERBLINK proper motion survey to 8980 square
degrees south of Decl.=-30$^{\circ}$. The new high proper motion stars
include 123 objects with $\mu>0.5\arcsec$ yr$^{-1}$, and 5 with
$\mu>1.0\arcsec yr^{-1}$. These new stars consist in a variety of
nearby red dwarfs and white dwarfs, and (slightly more distant) red
halo subdwarf, and are all prime targets for follow-up spectroscopic
and astrometric (parallax) observations. Comparison with previous
proper motion surveys in the southern sky suggests that SUPERBLINK has
a recovery rate between $80\%$ and $90\%$ at southern declinations
for stars with red magnitude $10<R_F<19$ and proper motion in the
range $0.5<\mu<2.0\arcsec$ yr$^{-1}$. This survey makes a significant
addition to the census of high proper motion stars at southern
declinations.
\end{abstract}



\section{Introduction}

Surveys of stars with large proper motions are very effective in
identifying low-luminosity objects in the vicinity of the Sun. Most of
the white dwarfs and red dwarfs in the Solar neighborhood have first
been discovered in proper motion surveys. Much of the current census of
nearby stars is based on follow-up analysis of objects listed in large
catalogs of high proper motion stars, most notably the NLTT
catalog of stars with $\mu>0.18\arcsec$ yr$^{-1}$ \citet{L79b}
and (especially) its subset, the LHS catalog of stars with
$\mu>0.5\arcsec$ yr$^{-1}$ \citet{L79a}. Objects with annual proper
motions in excess of $\mu=0.5\arcsec$ are now often referred to as
``LHS stars''. The current census of nearby stars is largely based on
follow-up observations of those Luyten objects
\citep{GJ91,GR97,Retal03}.

The Luyten catalogs have long been known \citep{D86} to suffer from
significant incompleteness both at low Galactic latitudes ($|b|<10$),
and in southernmost declinations ($Decl.<-32^{\circ}$). The desire
to build a complete census of stars in the Neighborhood of the Sun has
motivated a series of new proper motion surveys which, over the years,
have gradually filled up these gaps. Recently, improved data-mining
techniques have been used to conduct a massive proper motion survey of
the northern sky, using data from the Digitized Sky Survey (DSS). In a
systematic analysis of DSS scans with the SUPERBLINK software, an
automated blink comparator, several hundred new $\mu>0.5\arcsec$
yr$^{-1}$ stars have been discovered in the northern sky, effectively
filling up the low Galactic latitude gap \citep{LSR02}, and also
significantly increasing the completeness of the proper motion census
at high Galactic latitude \citep{LSR03}. In fact, the new {\em
  LSPM-north catalog} of stars with proper motions $\mu>0.15\arcsec$
yr$^{-1}$ \citep{LS05}, which compiles the SUPERBLINK results, is now
superseding the Luyten catalogs for all areas north of the celestial
equator.

The southern sky has been very attractive for proper motion hunters in
recent years, because of the low completeness of the LHS catalog at
southern declinations. Several small proper motion surveys have been
carried, including the Cal\'an-ESO (CE) Proper Motion Survey
\citep{RWRG01}, the WT survey \citep{WT97,WC01}, and the APMPM survey
\citep{SIIJM00}, which have all made significant, additions to the zoo
of $\mu>0.5\arcsec$ yr$^{-1}$ proper motion stars, although limited in
area. Two larger proper motion surveys have recently been conducted
based on data from the SuperCOSMOS Sky Survey (SSS). The
Liverpool-Edinburgh high proper motion (LEHPM) survey \citep{PJH03}
has covered $\approx7,000$ square degrees of the south Galactic cap,
including a large fraction of the largely incomplete
$Decl.<-32^{\circ}$ cap. The SuperCOSMOS-RECONS survey (SCR) now aims
to obtain an improved proper motion survey of the entire southern sky,
although results have so far been presented for $\approx5,000$
sq.deg. south of $Decl.<-47^{\circ}$ \citep{SHHBJ05}.

Encouraged by the impressive results already obtained with SUPERBLINK
in the northern sky, I have now expanded the analysis of DSS images to
the southern declinations. In this paper, I report the discovery of
a first set of 182 new stars with $\mu>0.45\arcsec$ yr$^{-1}$, found
after completing the analysis of $87.1\%$ of the sky south of
$Decl.<-30^{\circ}$ (8980 square degrees). The search methodology is
outlined in \S2. Results are described in \S3. The completeness of our
survey is estimated in \S4. Finder charts for all new objects are
provided in the appendix.


\section{Search and Identification}

Our method for identifying high proper motion stars from the Digitized
Sky Survey (DSS) is described in much detail in \citet{LS05}. Scans
from the DSS are analyzed using SUPERBLINK. For each detection,
SUPERBLINK generates two-epoch finding charts which are blinked on the
computer screen and examined by eye, in order to filter out any false
detection. The whole northern sky has already been analyzed with
SUPERBLINK, and the first major data release can be found in
\citet{LS05}.

Analysis of the southern declinations of the DSS by SUPERBLINK
presents an additional complication. In the northern hemisphere, the
DSS incorporates scans from photographic plates in the red passband at
both the first (POSS-I red; 103aE emulsion) and second epoch (POSS-II
red; IIIaF emulsion). In the south, however, the DSS only contains
first epoch blue plate (SERC-J blue; IIIaJ emulsion) while the second
epoch images are available only from red (SES red; IIIaF emulsion) and
far red plates (IVn emulsion). This is problematic, because SUPERBLINK
is based on image subtraction algorithms, and thus works best on pairs
of images that are as similar as possible. Nevertheless, tests have
shown the procedure to work reasonably well on pairs of images in
different passbands, as long as there is a sufficient number of stars
with similar plate intensities on both images. Tests have shown that
SUPERBLINK can recover high proper motion stars from pairs of blue/red
plates, although it fails to detect objects with extreme
colors. Encouraged by these results, it has been decided to expand the
SUPERBLINK analysis of the DSS to southern declinations. The analysis
of areas south of Decl.=-30$^{\circ}$ has now already been completed.

\begin{figure}
\epsscale{1.2}
\plotone{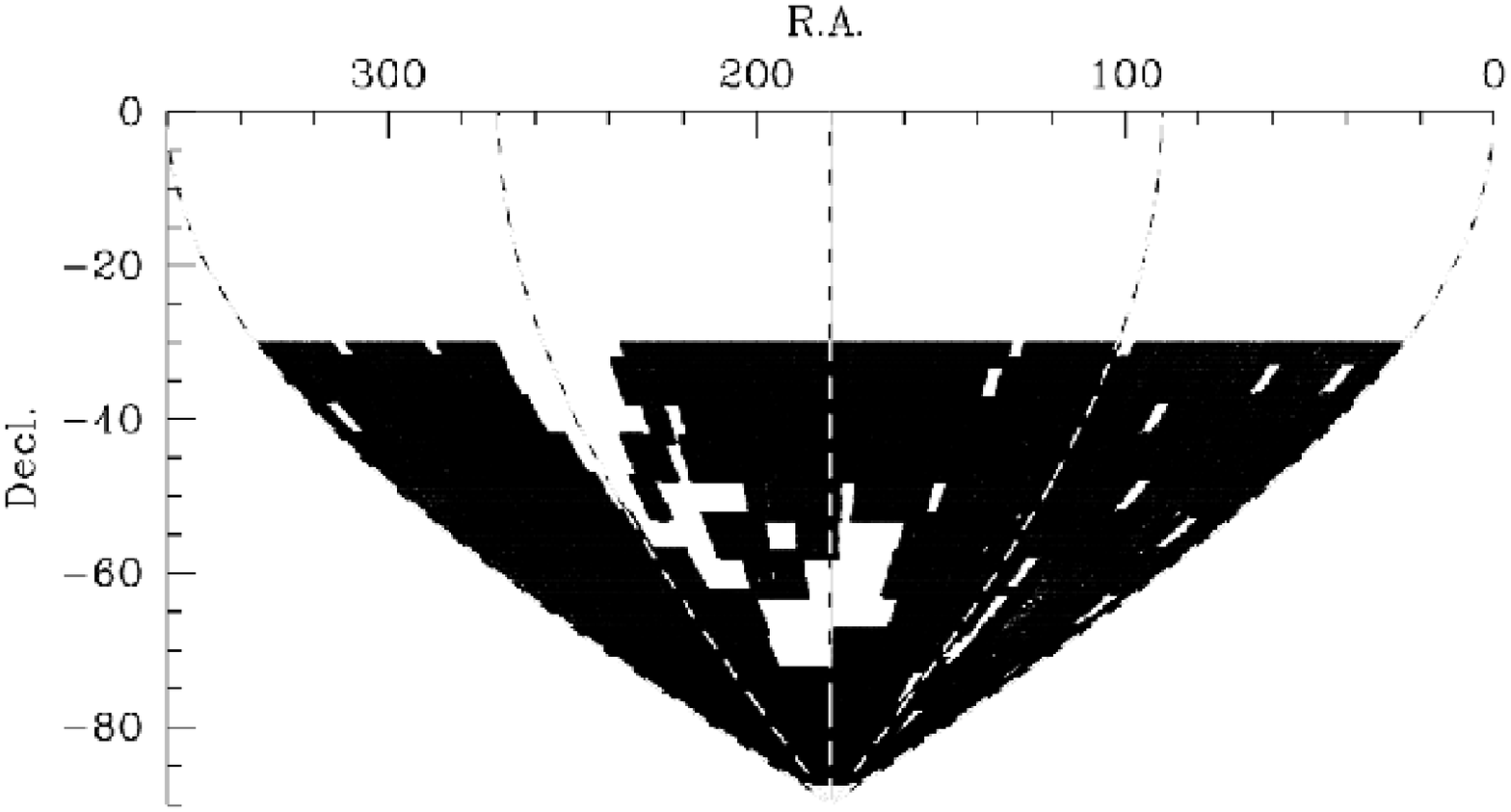}
\caption{Area currently covered in our southern sky proper motion
survey (shaded in black). The survey covers 8,980 square degrees, or
$87.1\%$ of the sky south of Decl.=-30$^{\circ}$. Several areas are
excluded from the survey because the DSS does not have a suitable
first epoch (in particular: several low Galactic latitude
plates). Other areas are excluded because the temporal baseline
between the first and second epoch images is too short.}
\end{figure}

Not all areas of the DSS are suitable for analysis by SUPERBLINK,
however. There are several zones where first epoch SERC-J plates are
unavailable, most notably at low Galactic latitudes. Proper motion
searches also require that there be a sufficient temporal baseline
between the first and second epoch. This was rather apparent in the
northern sky, where the POSS-I plates preceed the POSS-II plates by
$\approx40$ years. But in the southern sky, the baseline between the
SERC and SES surveys is significantly shorter ($<20$ years). In some
cases, the baseline is too short for accurate proper motion
detection/measurement. Pairs of plates with baselines under 5 years
were thus not considered for our analysis. The SUPERBLINK algorithms
also fail in subfields ($17\arcmin\times17\arcmin$) where there is an
extended, saturated object, such as very bright $V<5$ star. In the
end, 8980 square degrees, or $\simeq87.1\%$ of the sky south of
Decl.=-30$^{\circ}$, were successfully analyzed (Figure 1).

DSS scans were provided as input to the SUPERBLINK software. Lists of
candidate high proper motion stars were generated on output, along
with the usual two-epoch $4.25\arcmin\times4.25\arcmin$ finder
charts. The charts were visually examined by blinking on a computer
screen, and bogus detections were filtered out. Counterparts for all
confirmed objects were searched for in the {\em USNO-B1.0 catalog}
\citep{Metal03} and in the {\em 2MASS All-sky Catalog of Point Sources}
\citep{C03}. The 2MASS positions were used as a reference to calculate
the 2000.0 epoch position. Optical magnitudes were obtained from
USNO-B1.0, infrared magnitudes from 2MASS.


\section{High proper motion discoveries}

A total of 1,147 stars with proper motions in the range
$0.45<\mu<2.0\arcsec$ yr$^{-1}$ were located by SUPERBLINK. A majority
of these have been previously reported in the literature as high
proper motion stars. The list of SUPERBLINK detections was
cross-correlated with the NLTT catalog, and with lists of detections
from the APMPM survey \citep{SIIJM00}, the CE survey \citep{RWRG01},
the LEHPM catalog \citep{PJH03}, and the recent SCR proper motion
survey \citep{SHHBJ05}. A search was also made with
Simbad\footnote{http://simbad.u-strasbg.fr/}, to identify known high
proper motion stars from other sources, especially the WT stars
\citep{WT97,WC01}. In all, 965 of the SUPERBLINK detections were found
to be known high proper motion stars.

A total of 182 new high proper motion stars were discovered with
SUPERBLINK. A complete list is presented in Table 1. Names for the
stars are based on the convention introduced by \citet{E79,E80}. We
use the letters ``PM'' for {\em Proper Motion}, followed by a space,
followed by the letter ``J'' which denotes that the following digits
refer to the J2000 coordinates. The numerals then correspond to the
right ascension (to 0.1 minutes of time) and declination (to 1 minute
of arc). In case of confusion between two objects, most notably close
proper motion doubles, we add the directional suffix ``N'', ``S'',
``W'', or ``E'' to distinguish between the two components.

Finder charts for all the stars can be found in the appendix. Stars
range in magnitude from $R_F=11.3$ to $R_F=20.5$. Most notably, the
list includes 5 new stars with proper motions exceeding $1.0\arcsec$
yr$^{-1}$. These are PM J00522-6201, LPM J12277-4541, LPM J13384-3752,
LPM J19167-3638, and LPM J19261-4310. the list also includes two
common proper motion doubles: LPM J19105-4132 / LPM J19105-4133, and
LPM J22403-4931E / LPM J22403-4931W.

\begin{figure*}
\epsscale{1.1}
\plotone{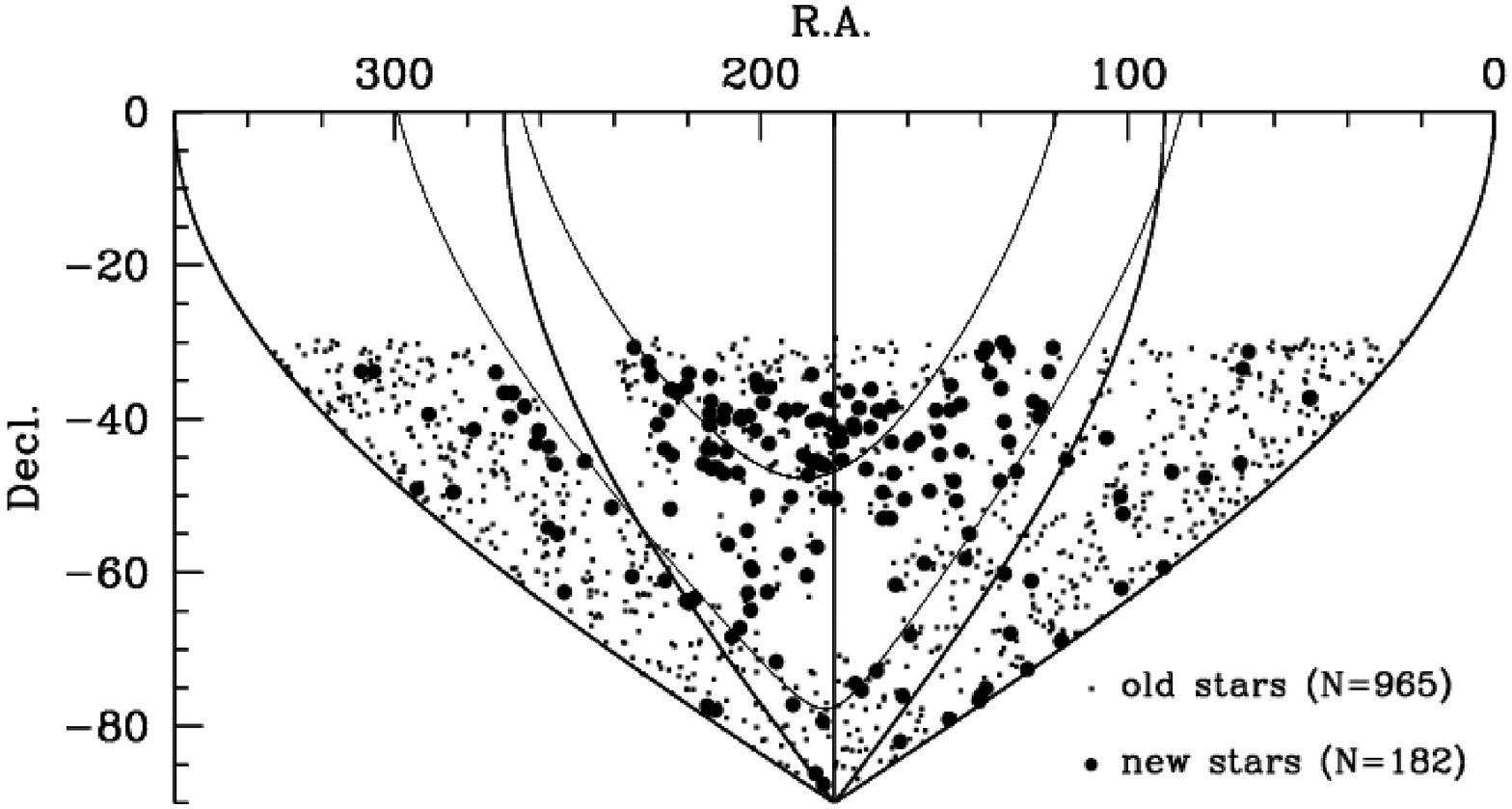}
\caption{Objects with $\mu>0.45\arcsec$ found in the survey area with
SUPERBLINK. The software correctly re-identified a total of 965
previously known high proper motion objects (``old
stars''). SUPERBLINK also discovered 182 new high proper motion
objects (``new stars''), which are here reported as high proper motion
stars for the first time.}
\end{figure*}

Figure 2 plots the distribution on the sky of all the stars found with
SUPERBLINK, with new discoveries denoted by larger symbols. New stars
are found all over the survey area, although relatively fewer ones are
found near the south Galactic cap (an area covered by the LEHPM
catalog), and in areas south of Decl.=-47$^{\circ}$ (where the SCR
survey has already made an efficient pass).

Table 1 provides J2000 coordinates for the stars at the 2000.0
epoch. The temporal baseline between the first and second epoch of the
discovery plates is given. The measured proper motions are {\em
  relative}, calculated with respect to background Galactic
stars. Proper motion errors ($\mu_{err}$) are estimated based on the
detailed analysis of SUPERBLINK errors from \citet{LS05}, with
normalization to the temporal baseline (a shorter baseline makes the
proper motion less accurate). Optical magnitudes B$_{\rm J}$, R$_{\rm
  F}$, and I$_{\rm   N}$ (respectively photographic IIIa-J, IIIa-F,
and IV-N) are listed; these are obtained from counterparts of the
stars found in the USNO-B1.0 catalog \citep{Metal03}. Infrared J, H,
and K$_s$ magnitudes are also given for stars that have counterparts
in the 2MASS All-Sky Catalog of Point Sources \citep{C03}.

\begin{figure}
\epsscale{1.25}
\plotone{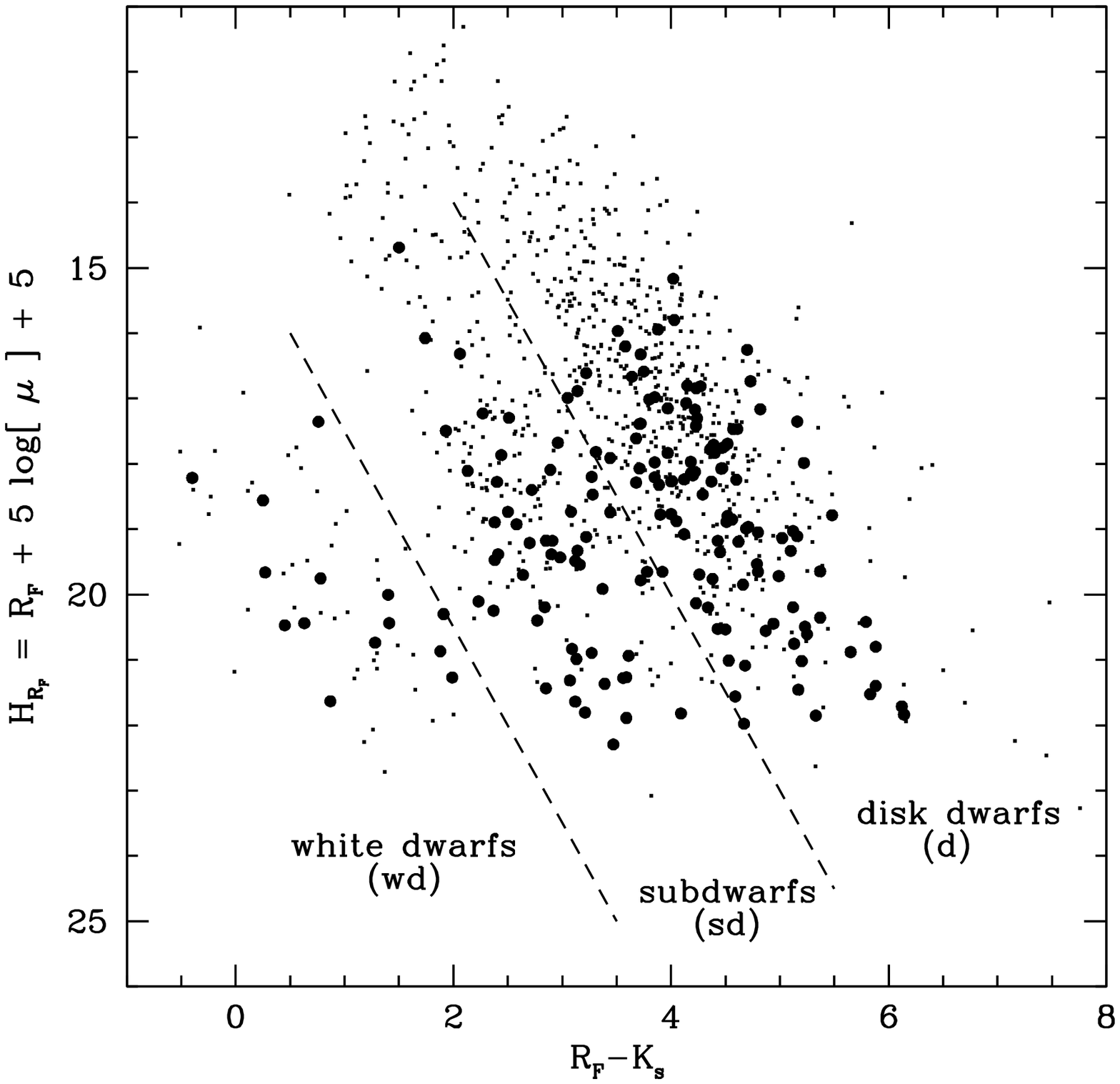}
\caption{Reduced proper motion diagram of the high proper motion stars
identified with SUPERBLINK. Small circles denote previously known
objects, and large circles denote new discoveries, as in Figure 2. The
approximate loci of white dwarfs, halo subdwarfs, and disk dwarfs are
indicated, as from \citet{LSR02}. The new SUPERBLINK stars tend to be
relatively faint ($R_F>14$) and so populate mostly the lower half of the
diagrams, where low-luminosity objects normally lie (white dwarfs,
low-mass dwarfs and subdwarfs). The new stars make a significant
addition to the census of objects with $H_{R_F}>20$.}
\end{figure}

A simple reduced proper motion analysis provides an estimate of the
spectral class of the stars. Optical to infrared proper motion
diagrams \citep{SG02} can be used to separate faint high proper motion
stars into three general classes: white dwarfs (wd), subdwarfs (sd),
and disk dwarfs (d). Here we use a $[H_{R_F},R_F-K_s]$ reduced
proper motion diagram and relationships defined in \citet{LRS03},
where $H_{R_F} = R_F + 5 log(\mu) + 5$. Figure 3 shows the
distribution, in the reduced proper motion diagram, of all 1,147 stars
found by SUPERBLINK, with the new discoveries again labeled with
larger symbols. Among the new discoveries, there are 14 candidate
white dwarfs, 52 candidate halo subdwarfs, and 106 candidate disk
dwarfs. The status of nine other stars remains ambiguous because of
incomplete photometric information.


\section{Completeness of SUPERBLINK in the southern sky}

To estimate the completeness of our survey, we have compared our list
of SUPERBLINK detections to lists of high proper motion stars from
five previous southern surveys which overlap with our search areas. 
Within the 8980 square degrees area analyzed with SUPERBLINK, we
compiled a total 769 stars with magnitude $10<R_F<19$ and proper
motion with $0.5<\mu<2.0\arcsec$ from the NLTT catalog \citep{L79b},
the APMPM survey \citep{SIIJM00}, the CE survey \citep{RWRG01}, the
LEHPM catalog \citep{PJH03}, and the SCR survey
\citep{{SHHBJ05}}. Note that stars brighter than magnitude
$R_F\approx10$ are generally not recovered by SUPERBLINK because they
form extended saturated patches on the SERC-J and SES plates; these are
thus excluded from the completeness analysis. Proper motions for all
those bright stars can be obtained from the TYCHO-2 catalog
\citep{H00}. From the reference sample of fainter ($R_F>10$) stars
SUPERBLINK recovered 612 objects, missing 157, suggesting recovery
rate of $79.5\%$.

It must be noted, however, that NLTT catalog entries are sometimes
affected by large positional errors \citep{SG03}. It is thus very
possible that a number of NLTT objects have been ``missed'' by
SUPERBLINK simply because they do not exist at the position recorded
in the NLTT catalog. Likewise, \citet{SHHBJ05}  report problems
recovering some the stars listed in the LEHPM catalog; it appears that
a number of LEHPM entries are either spurious, or have large errors
in their reported position. Further investigations would be required to
determine which of the NLTT and/or LEHPM objects are not real. In any
case, the SUPERBLINK recovery rate quoted above is probably an
underestimate. In fact, if we exclude from the analysis stars all the
stars that are listed only in the NLTT or LEHPM, we reduce the
reference sample to 259 previously known stars in our survey area, of
which only 26 have been missed by SUPERBLINK. This suggests a higher
recovery rate of $\approx90\%$. The actual recovery rate of SUPERBLINK
most probably lies somewhere between these two estimates.

In any case, it appears that the SUPERBLINK recovery rate in the south
is significantly lower than the much superior detection rate
($\approx99\%$) achieved at northern declinations \citep{LS05}. The
lower efficiency of SUPERBLINK in the southern sky is most likely due
to the fact that plates of different colors (blue/red) are used for
the first and second epoch. Because SUPERBLINK is an image-subtraction
analysis software, more confusion arises when plates of different
passbands are compared. In particular, one of our algorithms for the
identification of moving objects requires that they display
similar intensities (to within a factor of 2) on both the first and
second epoch plates.  This filter is required to minimize the number
of false detections from the often noisy/grainy photographic
plates. The software does renormalize the intensities so that the
total plate flux from all stars in the field are equal in the first
and second epoch image. In principle, if all the stars in the field
have the same color index, the renormalization will correct for it,
and the stars will show the same (renormalized) flux on both
images. However, if there are objects on the field that are bluer, or
redder than the typical background star, their renormalized flux will
be different on both images. The median value for the color of a field
star is $B_J-R_F\approx1.2$. In a typical field, where the red and
blue plates have been renormalized by such that a $B_J-R_F=1.2$ star
yields similar instrumental flux values on both images, stars with
$B_J-R_F<0.4$ or with $B_J-R_F>2.0$ will be missed by SUPERBLINK. In
practice, this effect is mitigated by the fact that the photographic
plate response is non-linear, but it remains true that unusually blue
or red stars are likely to be missed by the code. In fact, close
examination of some APMPM and SCR stars missed by SUPERBLINK shows
that they are indeed quite red. Preliminary tests reveal that some
tweaking of the SUPERBLINK software parameters, and relaxing of the
automated rejection filters would increase the efficiency of the code,
albeit at the expense of a larger number of false detections. These
would then have to be filtered out by eye, requiring a larger
investment of person-hours.


\section{Conclusions}

A search for high proper motion stars in the Digitized Sky Survey with
the SUPERBLINK software has been completed for an area of 8980 square
degrees covering $87.1\%$ of the sky south of $Decl.=-30^{\circ}$. A
total of 1,147 stars with proper motion $\mu>0.45\arcsec$ yr$^{-1}$
were detected by SUPERBLINK, of which 182 are found to be new
discoveries. Of these, there are 123 new stars with $\mu>0.5\arcsec$
yr$^{-1}$ (the canonical limit of the LHS catalog), including
5 stars with extremely large proper motions ($\mu>1.0\arcsec$
yr$^{-1}$). These make significant additions to the census of high
proper motion stars at southern declinations. 

A standard designation system is used to name the new high proper
motion stars. It is based on the naming convention first introduced by
\citet{E79,E80}. At the time, the standard system was used by Eggen as 
a means to simplify the designation of some high proper motion stars 
which were then known under a variety of names. The original
convention used PM{\it HHMMM${\pm}$ddmm}, for a star with R1950
coordinates R.A.={\it HH MM.M}, Decl.={\it${\pm}$dd mm}. Nowadays,
there has been several new surveys and catalogs, each using its own
prefix system, with much confusion resulting. We are again
reaching a point where it is becoming desirable to uniformize the
naming convention of the high proper motion stars. I propose a slight
modification to the Eggen convention, in which the J2000 coordinates
are used instead. High proper motion stars are thus to be designated
according to the model PM J{\it HHMMM${\pm}$ddmm}, for a star at the
J2000 coordinates R.A.={\it HH MM.M} Decl.={\it${\pm}$dd mm}. I
strongly encourage all researchers in the field to start following
this convention as well, at least when naming a new discovery.
 
A reduced proper motion analysis suggests that 14 of the new high
proper motion stars are white dwarfs, while 52 more are candidate
halo subdwarfs. Most of the others are probably nearby red
dwarfs. All these objects should be high priority targets for follow-up
spectroscopy and astrometry (parallax measurements).

One notes that in the area covered by our survey, there are now 940
known $R_F>10$ stars with $0.5<\mu<2.0\arcsec$ yr$^{-1}$, where
Luyten's NLTT catalog listed 564. Much of the improvement comes from
the fact that the SES blue and AAO red plates used in the
SUPERBLINK survey are significantly deeper (20th magnitude) than the
Bruce Proper Motion survey plates used by Luyten in the south, which
had a limiting photographic magnitude $m_{pg}\approx15.5$.

At southern declinations, the SUPERBLINK software has a detection rate
of $80\%-90\%$ for stars with $10<R_F<19$, which falls short of
matching the very high detection rates $\approx99\%$ achieved in the
northern sky survey. The main reason is that in the south, the DSS has
only first epoch scans of blue (IIIaJ) plates, but second epoch scans
of red (IIIaF) plates, while in the north, red plates were available
for both the first and second epochs. The image subtraction techniques
used by SUPERBLINK makes handling of blue/red pairs more
problematic. Strategies to increase the SUPERBLINK detection rate of
high proper motion stars in pairs of blue/red images are currently
being considered and tested. In the meantime, the SUPERBLINK analysis
of the remaining half of the southern sky
($0^{\circ}<Decl.<-30^{\circ}$) is now being performed, and new
SUPERBLINK discoveries will be available soon.


\acknowledgments

{\bf Acknowledgments}

The author gratefully acknowledges support from the Cordelia
Corporation, from Hilary Lipsitz, and from the American Museum of
Natural History.

This work has been made possible through the use of the Digitized Sky
Surveys. The Digitized Sky Surveys were produced at the Space
Telescope Science Institute under U.S. Government grant NAG
W-2166. The images of these surveys are based on photographic data
obtained using the Oschin Schmidt Telescope on Palomar Mountain and
the UK Schmidt Telescope. The plates were processed into the present
compressed digital form with the permission of these institutions. The
National Geographic Society - Palomar Observatory Sky Atlas (POSS-I)
was made by the California Institute of Technology with grants from
the National Geographic Society. The Second Palomar Observatory Sky
Survey (POSS-II) was made by the California Institute of Technology
with funds from the National Science Foundation, the National
Geographic Society, the Sloan Foundation, the Samuel Oschin
Foundation, and the Eastman Kodak Corporation. The Oschin Schmidt
Telescope is operated by the California Institute of Technology and
Palomar Observatory. The UK Schmidt Telescope was operated by the
Royal Observatory Edinburgh, with funding from the UK Science and
Engineering Research Council (later the UK Particle Physics and
Astronomy Research Council), until 1988 June, and thereafter by the
Anglo-Australian Observatory. The blue plates of the southern Sky
Atlas and its Equatorial Extension (together known as the SERC-J), as
well as the Equatorial Red (ER), and the Second Epoch [red] Survey
(SES) were all taken with the UK Schmidt. 

This publication makes use of data products from the Two Micron All
Sky Survey, which is a joint project of the University of
Massachusetts and the Infrared Processing and Analysis
Center/California Institute of Technology, funded by the National
Aeronautics and Space Administration and the National Science
Foundation.

The data mining required for this work has been made possible with the
use of the SIMBAD astronomical database and VIZIER astronomical
catalogs service, both maintained and operated by the Centre de
Donn\'ees Astronomiques de Strasbourg (http://cdsweb.u-strasbg.fr/).

\newpage



\appendix

\section{Finder charts}

Finder charts of high proper motion stars are generated as a
by-product of the SUPERBLINK software. We present here finder charts
for all the new, high proper motion stars presented in this paper and
listed in Table 1. All the charts consist of pairs of
$4.25\arcmin\times4.25\arcmin$ images showing the local field at two
different epochs. The name of the star is indicated in the center just
below the chart, and corresponds to the name given in Table 1. To the
left is the POSS-I field, with the epoch of the field noted in the
lower left corner. To the right is the modified POSS-II field which
has been shifted, rotated, and degraded in such a way that it matches
the quality and aspect of the POSS-I image. The epoch of the POSS-II
field is noted on the lower right corner. High proper motion stars are
identified with circles centered on their positions at the epoch on
the plate.

The charts are oriented in the local X-Y coordinate system of the
POSS-I image; the POSS-II image has been remapped on the POSS-I
grid. This means that north is generally up and east left, but the
fields might appear rotated by a small angle for high declination
objects. Sometimes a part of the field is missing: this is an artifact
of the code. SUPERBLINK works on $17\arcmin\times17\arcmin$ DSS
subfields. If a high proper motion star is identified near the edge of
that subfield, the output chart appears truncated.


\begin{figure}
\epsscale{1.2}
\caption{Finding charts for the new high proper motion stars
discovered with SUPERBLINK, as listed in Table 1.}
\end{figure}

\begin{figure}
\epsscale{1.2}
\caption{Finding charts for the new high proper motion stars
discovered with SUPERBLINK (continued).}
\end{figure}

\begin{figure}
\epsscale{1.2}
\caption{Finding charts for the new high proper motion stars
discovered with SUPERBLINK (continued).}
\end{figure}

\begin{figure}
\epsscale{1.2}
\caption{Finding charts for the new high proper motion stars
discovered with SUPERBLINK (continued).}
\end{figure}

\begin{figure}
\epsscale{1.2}
\caption{Finding charts for the new high proper motion stars
discovered with SUPERBLINK (continued).}
\end{figure}

\begin{figure}
\epsscale{1.2}
\caption{Finding charts for the new high proper motion stars
discovered with SUPERBLINK (continued).}
\end{figure}

\begin{figure}
\epsscale{1.2}
\caption{Finding charts for the new high proper motion stars
discovered with SUPERBLINK (continued).}
\end{figure}

\begin{figure}
\epsscale{1.2}
\caption{Finding charts for the new high proper motion stars
discovered with SUPERBLINK (continued).}
\end{figure}

\begin{figure}
\epsscale{1.2}
\caption{Finding charts for the new high proper motion stars
discovered with SUPERBLINK (continued).}
\end{figure}

\clearpage
\LongTables
\begin{landscape}
\begin{deluxetable}{lrrrrrrrrrrrrrr}
\tabletypesize{\scriptsize}
\tablecolumns{13}
\tablewidth{0pt}
\tablecaption{New High proper motion stars.}
\tablehead{
\colhead{Star} & 
\colhead{$\alpha$(J2000)} &
\colhead{$\delta$(J2000)} &
\colhead{$\mu$} & 
\colhead{$\mu_{RA}$} & 
\colhead{$\mu_{Decl}$} & 
\colhead{$\Delta t$\tablenotemark{a}} & 
\colhead{$\mu_{err}$\tablenotemark{b}} & 
\colhead{$B_J$\tablenotemark{c}} & 
\colhead{$R_F$} &
\colhead{$I_N$} & 
\colhead{$J$\tablenotemark{d}} &
\colhead{$H$} &
\colhead{$K_s$}&
\colhead{class}\\
\colhead{} & 
\colhead{} &
\colhead{} &
\colhead{($\arcsec$ yr$^{-1}$)} & 
\colhead{($\arcsec$ yr$^{-1}$)} & 
\colhead{($\arcsec$ yr$^{-1}$)} & 
\colhead{(yr)} & 
\colhead{($\arcsec$ yr$^{-1}$)} & 
\colhead{} & 
\colhead{} &
\colhead{} & 
\colhead{} &
\colhead{} &
\colhead{} &
\colhead{}
}
\startdata 

PM J00151-5919 &  0 15 10.13& -59 19 56.7&   0.483&  0.473&  0.097& 16.02& 0.020&   19.6& 17.0& 13.5& 12.14& 11.56& 11.21             &  d\\
PM J00168-7233 &  0 16 51.97& -72 33 46.4&   0.507& -0.383& -0.332&  9.92& 0.032&   19.0& 17.0& 14.8& 13.33& 12.84& 12.57             &  d\\
PM J00294-6900 &  0 29 26.87& -69 00 13.3&   0.486&  0.465& -0.139& 13.89& 0.023&   18.0& 16.7& 14.3& 13.22& 12.74& 12.47             &  d\\
PM J00327-7641 &  0 32 43.62& -76 41 28.7&   0.593&  0.592&  0.032&  9.92& 0.032&   19.3& 17.4& 15.5& 14.58& 14.03& 13.81             & sd\\
PM J00344-6849 &  0 34 28.77& -68 49 29.7&   0.508& -0.016& -0.508& 13.89& 0.023&   19.5& 18.8& 17.8& \nodata& \nodata& \nodata       &  \nodata\\
PM J00522-6201 &  0 52 15.28& -62 01 54.6&   1.073&  1.065&  0.122& 15.09& 0.021&   19.7& 16.7& 13.3& 12.15& 11.74& 11.37             &  d\\
PM J00568-7603 &  0 56 53.20& -76 03 43.8&   0.457&  0.194& -0.414& 13.96& 0.023&\nodata& 17.5& 15.1& 12.57& 11.99& 11.62          &  d\\
PM J01009-7904 &  1 00 56.07& -79 04 25.2&   0.450& -0.270& -0.360& 14.97& 0.021&\nodata& 11.9& 10.0&  8.80&  8.17&  7.88          &  d\\
PM J01082-3714W&  1 08 13.21& -37 14 39.2&   0.509& -0.128& -0.492& 17.39& 0.018&\nodata& 15.2& 13.2& 12.84& 12.36& 12.12          & sd\\
PM J01082-3714E&  1 08 13.69& -37 14 39.1&   0.509& -0.128& -0.492& 17.39& 0.018&\nodata& 15.2& 14.1& 13.41& 12.97& 12.70          & sd\\
PM J01184-7504 &  1 18 29.66& -75 04 55.0&   0.454&  0.446&  0.081& 13.96& 0.023&   19.7& 15.5& \nodata& 10.85& 10.29& 10.02          &  d\\
PM J01253-4545 &  1 25 18.02& -45 45 31.1&   0.748&  0.517& -0.541& 13.82& 0.023&   18.0& 16.9& 18.7& 15.11& 14.84& 14.91             & wd\\
PM J01598-4736 &  1 59 48.21& -47 36 12.3&   0.466&  0.272& -0.378& 13.05& 0.025&\nodata& 16.2& 15.1& 13.80& 13.30& 13.04          & sd\\
PM J03007-4653 &  3 00 45.15& -46 53 50.6&   0.745&  0.676&  0.312& 16.95& 0.019&   18.3& 15.4& 12.7& 11.79& 11.31& 11.02             &  d\\
PM J03064-3325 &  3 06 26.46& -33 25 42.9&   0.972& -0.164& -0.957& 16.01& 0.020&   19.5& 16.7& 15.8& 14.35& 13.85& 13.58             & sd\\
PM J03116-3113 &  3 11 37.99& -31 13 21.2&   0.544&  0.162& -0.520& 13.99& 0.023&   20.8& 18.3& 16.3& 14.37& 13.88& 13.63             &  d\\
PM J03229-8159 &  3 22 59.54& -81 59 54.4&   0.504&  0.383&  0.327& 12.11& 0.026&\nodata& 18.2& 14.8& 13.02& 12.48& 12.08          &  d\\
PM J03245-5220 &  3 24 32.99& -52 20 39.3&   0.823& -0.444& -0.692& 11.92& 0.027&   17.2& 14.6& 11.6& 11.10& 10.51& 10.17             &  d\\
PM J03286-6756 &  3 28 38.30& -67 56 26.5&   0.469&  0.451&  0.129&  5.28& 0.061&\nodata& 17.2& 14.7& 13.09& 12.66& 12.33          &  d\\
PM J03536-5006 &  3 53 36.57& -50 06 00.9&   0.523&  0.255&  0.457& 17.21& 0.019&\nodata& 20.5& 18.5& \nodata& \nodata& \nodata    &  \nodata\\
PM J04356-6105 &  4 35 37.39& -61 05 40.4&   0.502&  0.432&  0.256& 14.71& 0.022&   17.5& 16.5& 16.0& 15.70& 15.21& 15.10             & wd\\
PM J05175-4227 &  5 17 35.96& -42 27 54.4&   0.452&  0.191& -0.410& 15.06& 0.021&\nodata& 12.8& 11.7& 11.80& 11.31& 11.06          & sd\\
PM J05472-6009 &  5 47 14.62& -60 09 50.5&   0.475&  0.164& -0.445& 17.28& 0.019&   20.9& 16.4& \nodata& 13.39& 12.89& 12.68          & sd\\
PM J05597-4519 &  5 59 44.05& -45 19 10.5&   0.819&  0.449&  0.685& 12.87& 0.025&   18.7& 16.8& 15.4& 14.15& 13.61& 13.41             & sd\\
PM J06491-7603 &  6 49 06.65& -76 03 50.1&   0.512&  0.428&  0.280& 12.88& 0.025&\nodata& 16.7& 16.0& 14.88& 14.43& 14.33          & sd\\
PM J07096-3941 &  7 09 37.05& -39 41 52.0&   0.472& -0.124& -0.455& 14.86& 0.022&   16.7& 14.7& 12.5& 11.77& 11.21& 10.99             &  d\\
PM J07096-4648 &  7 09 37.27& -46 48 59.1&   0.456&  0.137&  0.434& 14.81& 0.022&\nodata& 14.0& 13.1& 12.20& 11.70& 11.49          & sd\\
PM J07110-3824 &  7 11 00.96& -38 24 46.1&   0.893&  0.516& -0.728& 14.06& 0.023&   20.6& 15.6& 13.0& 11.09& 10.55& 10.23             &  d\\
PM J07185-3351 &  7 18 30.75& -33 51 24.5&   0.461& -0.024&  0.461& 14.93& 0.021&   19.0& 18.4& 17.8& \nodata& \nodata& \nodata       &  \nodata\\
PM J07228-3042 &  7 22 51.34& -30 42 34.4&   0.494&  0.323&  0.374& 15.11& 0.021&   18.3& 17.4& 17.2& 16.42& 15.69& 15.52             & wd\\
PM J07253-3742 &  7 25 21.57& -37 42 59.1&   0.587&  0.211& -0.548& 14.06& 0.023&   13.3& 12.1& 10.1&  9.06&  8.42&  8.22             &  d\\
PM J07283-5812 &  7 28 23.20& -58 12 31.1&   0.482&  0.367& -0.312& 14.83& 0.022&\nodata& 12.9& 12.4& 11.60& 10.97& 10.84          & sd\\
PM J07293-4804 &  7 29 21.66& -48 04 27.3&   0.450&  0.361&  0.266& 14.16& 0.023&   14.5& 12.7& 10.9& 10.02&  9.44&  9.19             &  d\\
PM J07401-4257 &  7 40 11.83& -42 57 40.7&   0.703& -0.436&  0.551& 14.00& 0.023&   15.5& 12.5& 10.3&  8.68&  8.09&  7.77             &  d\\
PM J07433-5457 &  7 43 18.57& -54 57 30.0&   0.484& -0.098&  0.474& 14.89& 0.021&\nodata& 18.1& 15.1& 13.19& 12.61& 12.27          &  d\\
PM J07571-4021 &  7 57 07.87& -40 21 33.5&   0.551& -0.532&  0.143& 12.20& 0.026&   17.3& 14.5& 15.7& 11.41& 10.92& 10.65             &  d\\
PM J08152-3600 &  8 15 15.99& -36 00 59.5&   0.639& -0.105&  0.630& 15.17& 0.021&   16.2& 15.0& 12.6& 10.74& 10.17&  9.88             &  d\\
PM J08172-6808 &  8 17 16.66& -68 08 36.7&   0.551& -0.301&  0.461& 14.06& 0.023&   18.9& 16.8& 15.5& \nodata& \nodata& \nodata       &  \nodata\\
PM J08186-3110 &  8 18 40.24& -31 10 19.6&   0.817&  0.224& -0.785& 14.05& 0.023&   16.1& 15.1& 14.9& 14.92& 14.73& 14.83             & wd\\
PM J08276-3003 &  8 27 40.89& -30 03 00.0&   0.594& -0.326&  0.497& 14.05& 0.023&   16.3& 14.1& 12.2& 10.67& 10.17&  9.92             &  d\\
PM J08300-5039 &  8 30 00.78& -50 39 46.2&   0.989& -0.568&  0.810& 14.11& 0.023&   15.7& 13.8& 12.5& 10.70& 10.15&  9.90             &  d\\
PM J08355-3400 &  8 35 31.73& -34 00 36.9&   0.546& -0.138& -0.529& 12.26& 0.026&   16.0& 14.3& 11.2&  9.90&  9.37&  9.08             &  d\\
PM J08446-4805 &  8 44 38.92& -48 05 21.8&   0.748& -0.224&  0.713& 14.11& 0.023&   14.4& 12.7& 10.7&  9.36&  8.81&  8.56             &  d\\
PM J08458-3051 &  8 45 51.96& -30 51 31.3&   0.518& -0.477& -0.201& 13.97& 0.023&   16.4& 14.5& 12.7& 10.82& 10.30& 10.04             &  d\\
PM J08463-4407 &  8 46 20.99& -44 07 22.0&   0.477&  0.462& -0.117& 15.17& 0.021&   17.5& 16.3& 14.8& 12.80& 12.26& 12.04             &  d\\
PM J08471-3046 &  8 47 09.70& -30 46 12.0&   0.528&  0.062& -0.524& 13.97& 0.023&   15.5& 13.4& 12.1& 10.39&  9.91&  9.60             &  d\\
PM J08496-3138 &  8 49 38.94& -31 38 22.9&   0.506& -0.174&  0.475& 13.97& 0.023&   16.4& 14.8& 13.2& 11.69& 11.16& 10.91             &  d\\
PM J08503-5848 &  8 50 21.34& -58 48 06.7&   0.612& -0.418&  0.447& 14.84& 0.022&   18.0& 17.7& 17.1& 16.74& 16.90& 16.83             & wd\\
PM J09047-3804 &  9 04 46.49& -38 04 07.0&   0.614&  0.408& -0.458& 14.80& 0.022&   17.3& 14.8& 13.7& 12.03& 11.57& 11.36             & sd\\
PM J09172-3849 &  9 17 13.61& -38 49 36.0&   0.494& -0.086&  0.486& 14.80& 0.022&   21.4& 14.8& 13.5& 11.56& 11.12& 10.80             &  d\\
PM J09181-4437 &  9 18 08.62& -44 37 24.5&   0.501& -0.405&  0.295& 13.98& 0.023&   17.3& 16.8& 16.5& 15.59& 15.37& 14.89             & wd\\
PM J09204-4922 &  9 20 26.02& -49 22 34.7&   0.561& -0.406&  0.387& 14.81& 0.022&   17.8& 15.9& 12.7& 11.33& 10.82& 10.53             &  d\\
PM J09233-3537 &  9 23 23.68& -35 37 58.3&   0.485& -0.437& -0.210& 14.03& 0.023&   16.8& 13.8& 12.9& 12.24& 11.72& 11.53             & sd\\
PM J09237-7248 &  9 23 42.50& -72 48 36.7&   0.472& -0.360&  0.305& 14.88& 0.022&\nodata& 15.1& 13.6& 12.63& 12.07& 11.82          & sd\\
PM J09271-4137 &  9 27 07.19& -41 37 12.0&   0.476&  0.389& -0.274& 14.28& 0.022&\nodata& 11.3& 11.0& 10.32&  9.89&  9.80          & sd\\
PM J09376-3852 &  9 37 36.23& -38 52 23.0&   0.586& -0.446&  0.380& 14.28& 0.022&\nodata& 16.6& 16.5& 16.03& 15.60& 15.97          & wd\\
PM J09396-6135 &  9 39 38.82& -61 35 11.3&   0.481&  0.467&  0.112& 14.81& 0.022&   19.1& 13.2& \nodata& 10.80& 10.30&  9.98          &  d\\
PM J09566-4234 &  9 56 36.99& -42 34 27.3&   0.658&  0.382& -0.536& 14.99& 0.021&\nodata& 14.9& 12.5& 10.99& 10.47& 10.21          &  d\\
PM J09598-5027 &  9 59 53.94& -50 27 17.7&   0.677& -0.665&  0.128& 15.83& 0.020&   16.4& 15.6& 15.6& 15.46& 15.20& 14.82             & wd\\
PM J10026-7519 & 10 02 37.66& -75 19 22.7&   0.523& -0.371&  0.368& 14.95& 0.021&\nodata& 16.9& 14.6& 12.60& 11.98& 11.67          &  d\\
PM J10050-4322 & 10 05 03.11& -43 22 28.2&   0.632& -0.589&  0.228& 14.07& 0.023&   15.1& 13.3& 11.0&  9.85&  9.32&  9.06             &  d\\
PM J10185-5254 & 10 18 30.01& -52 54 44.3&   0.514& -0.297&  0.419& 19.10& 0.017&   16.4& 13.8& 13.3& 13.74& 13.29& 13.04             & wd\\
PM J10256-4703 & 10 25 39.40& -47 03 51.1&   0.477& -0.445&  0.170& 14.07& 0.023&   19.9& 17.5& 16.2& 15.05& 14.63& 14.23             & sd\\
PM J10326-5255 & 10 32 40.69& -52 55 05.8&   0.558& -0.512&  0.221& 19.10& 0.017&\nodata& 15.8& \nodata& 11.98& 11.38& 11.01       &  d\\
PM J10346-4259 & 10 34 37.86& -42 59 57.0&   0.471& -0.422&  0.208& 14.06& 0.023&   16.7& 14.9& 13.4& 11.65& 11.15& 10.89             &  d\\
PM J10352-7424 & 10 35 12.32& -74 24 57.0&   0.461& -0.397&  0.233& 14.95& 0.021&\nodata& 15.8& \nodata& 13.40& 12.84& 12.58       & sd\\
PM J10382-4933 & 10 38 15.18& -49 33 44.6&   0.682& -0.333& -0.595& 15.03& 0.021&\nodata& 13.9& 11.7& 10.31&  9.72&  9.43          &  d\\
PM J10395-3820 & 10 39 32.40& -38 20 02.0&   0.702& -0.196& -0.674& 13.40& 0.024&   16.9& 15.1& \nodata& 10.85& 10.30& 10.00          &  d\\
PM J10538-3858 & 10 53 49.42& -38 58 59.0&   0.558& -0.347&  0.436& 18.10& 0.018&   15.7& 14.1& 12.4& 10.91& 10.43& 10.13             &  d\\
PM J10587-3854 & 10 58 47.12& -38 54 14.8&   0.622& -0.603&  0.151& 18.10& 0.018&   16.4& 14.5& 12.4& 11.01& 10.52& 10.21             &  d\\
PM J11076-4105 & 11 07 40.05& -41 05 44.2&   0.492& -0.475&  0.129& 14.90& 0.021&   22.3& 18.1& 16.0& 14.28& 13.80& 13.51             &  d\\
PM J11094-4631 & 11 09 28.34& -46 31 10.0&   0.452& -0.447&  0.066& 18.77& 0.017&\nodata& 15.6& 14.0& 12.35& 11.88& 11.55          &  d\\
PM J11104-3608 & 11 10 29.04& -36 08 24.5&   0.535& -0.531& -0.071& 14.13& 0.023&\nodata& 14.6& 18.7& 10.93& 10.34& 10.00          &  d\\
PM J11256-3834 & 11 25 37.31& -38 34 42.9&   0.541& -0.528& -0.119& 14.90& 0.021&\nodata& 13.8& \nodata& 10.09&  9.51&  9.19       &  d\\
PM J11308-4120 & 11 30 48.27& -41 20 20.9&   0.454& -0.441& -0.107& 14.90& 0.021&   16.8& 15.0& 13.2& 12.09& 11.52& 11.32             &  d\\
PM J11329-4039 & 11 32 57.95& -40 39 21.6&   0.786& -0.685&  0.385& 19.14& 0.017&   15.7& 13.5& 11.5& 10.38&  9.89&  9.65             &  d\\
PM J11413-3624 & 11 41 21.53& -36 24 34.7&   0.590&  0.532&  0.255& 15.93& 0.020&\nodata& 12.4& 10.1&  8.49&  7.97&  7.70          &  d\\
PM J11480-4523 & 11 48 03.32& -45 23 01.8&   0.635& -0.536& -0.340& 14.12& 0.023&   16.9& 14.2& 13.8& 14.89& 14.76& 14.60             & wd\\
PM J11495-4248 & 11 49 31.69& -42 48 10.4&   0.989& -0.979& -0.136& 19.14& 0.017&   16.0& 13.3& 12.3& 11.67& 11.11& 10.90             & sd\\
PM J11511-4142 & 11 51 07.82& -41 42 17.3&   0.600& -0.569& -0.189& 19.14& 0.017&\nodata& 15.3& 12.8& 11.51& 10.99& 10.68          &  d\\
PM J11578-5022 & 11 57 50.47& -50 22 32.1&   0.478& -0.397& -0.265& 14.85& 0.022&\nodata& 12.4& 10.8&  9.27&  8.65&  8.37          &  d\\
PM J11596-4256 & 11 59 37.65& -42 56 39.1&   0.603& -0.446& -0.405& 17.65& 0.018&   15.4& 12.3& 10.6&  9.54&  8.98&  8.72             &  d\\
PM J12042-4037 & 12 04 15.50& -40 37 52.1&   0.603&  0.230& -0.557& 17.06& 0.019&   14.9& 12.9& 10.8&  9.57&  9.02&  8.75             &  d\\
PM J12088-3723 & 12 08 51.03& -37 23 27.3&   0.477&  0.346& -0.329& 15.86& 0.020&   16.0& 14.3& 12.0& 10.62& 10.08&  9.78             &  d\\
PM J12146-4603 & 12 14 40.02& -46 03 14.2&   0.797& -0.747& -0.276& 17.65& 0.018&   16.9& 14.6& 12.3& 10.32&  9.75&  9.44             &  d\\
PM J12159-5014 & 12 15 56.96& -50 14 19.3&   0.665& -0.665&  0.001& 14.85& 0.022&   19.0& 17.2& 15.8& 14.64& 14.21& 14.13             & sd\\
PM J12201-4546 & 12 20 08.01& -45 46 18.3&   0.750& -0.702&  0.262& 17.65& 0.018&\nodata& 14.8& 13.6& 12.70& 12.16& 11.95          & sd\\
PM J12225-4002 & 12 22 32.28& -40 02 12.2&   0.602& -0.577& -0.170& 17.06& 0.019&   19.8& 17.5& 14.5& 12.60& 11.98& 11.62             &  d\\
PM J12277-4541 & 12 27 46.91& -45 41 17.1&   1.321& -1.285&  0.301& 15.11& 0.021&   16.4& 14.5& 13.2& 12.75& 12.40& 12.27             & sd\\
PM J12300-3411 & 12 30 01.77& -34 11 23.9&   0.563& -0.465& -0.317& 15.92& 0.020&   15.3& 13.6& 11.2&  9.34&  8.77&  8.44             &  d\\
PM J12313-4018 & 12 31 21.19& -40 18 36.8&   0.569& -0.568& -0.023& 13.02& 0.025&   16.2& 14.0& 12.1& 10.44&  9.94&  9.64             &  d\\
PM J12346-5640 & 12 34 40.35& -56 40 12.4&   0.559& -0.527& -0.183& 14.65& 0.022&   19.8& 16.8& \nodata& \nodata& \nodata& \nodata    &  \nodata\\
PM J12355-4527 & 12 35 35.00& -45 27 03.8&   0.450& -0.302&  0.333& 15.11& 0.021&   15.3& 13.4& 11.9& 10.57& 10.04&  9.76             &  d\\
PM J12415-4717 & 12 41 33.18& -47 17 05.7&   0.471& -0.454& -0.125& 15.11& 0.021&   16.2& 14.5& 14.2& 12.77& 12.21& 12.06             & sd\\
PM J12472-4441 & 12 47 15.88& -44 41 50.0&   0.688& -0.683&  0.080& 15.11& 0.021&   19.0& 17.7& 16.2& 14.65& 14.26& 14.11             & sd\\
PM J12515-3846 & 12 51 31.82& -38 46 12.5&   0.625& -0.616&  0.102& 15.10& 0.021&   18.6& 16.9& 13.9& 12.09& 11.56& 11.25             &  d\\
PM J13002-6024 & 13 00 15.78& -60 24 19.3&   0.490& -0.437&  0.221& 15.02& 0.021&   16.9& 15.4& 12.9& 11.76& 11.21& 10.84             &  d\\
PM J13077-7925 & 13 07 44.45& -79 25 10.3&   0.514& -0.486& -0.165& 17.10& 0.019&\nodata& 17.8& 17.3& \nodata& \nodata& \nodata    &  \nodata\\
PM J13084-3903 & 13 08 27.01& -39 03 32.0&   0.456& -0.443& -0.110& 15.10& 0.021&\nodata& 16.9& 14.1& 12.60& 12.07& 11.78          &  d\\
PM J13095-3848 & 13 09 31.06& -38 48 04.4&   0.572& -0.554& -0.141& 15.10& 0.021&\nodata& 12.8& 10.9&  9.86&  9.30&  9.05          &  d\\
PM J13139-5011 & 13 13 54.14& -50 11 09.6&   0.468& -0.414& -0.216& 15.04& 0.021&   18.9& 16.3& 13.7& 12.38& 11.83& 11.50             &  d\\
PM J13276-3551 & 13 27 39.59& -35 51 01.0&   0.518& -0.404& -0.325& 15.94& 0.020&   16.8& 14.7& 12.3& 11.13& 10.60& 10.33             &  d\\
PM J13338-5738 & 13 33 53.03& -57 38 42.5&   0.710& -0.695& -0.144& 17.93& 0.018&   19.1& 17.2& 15.0& 12.84& 12.32& 12.03             &  d\\
PM J13379-4311 & 13 37 56.06& -43 11 30.1&   0.522& -0.408& -0.325& 16.03& 0.020&   17.1& 15.3& 13.5& 11.70& 11.11& 10.79             &  d\\
PM J13384-3752 & 13 38 26.53& -37 52 50.4&   1.265& -1.264& -0.031& 15.94& 0.020&\nodata& 15.5& 13.2& 11.75& 11.28& 10.97          &  d\\
PM J13423-3534 & 13 42 21.29& -35 34 50.8&   0.903& -0.878& -0.211& 15.94& 0.020&   19.0& 16.5& \nodata& 13.76& 13.22& 12.94          & sd\\
PM J13420-3544 & 13 42 00.21& -35 44 51.5&   0.468& -0.454&  0.113& 15.94& 0.020&   18.8& 16.3& \nodata& 13.31& 12.80& 12.52          & sd\\
PM J13438-3447 & 13 43 48.93& -34 47 49.5&   0.487& -0.479& -0.086& 15.94& 0.020&   18.5& 17.0& \nodata& 16.25& 15.90& 15.59          & wd\\
PM J13548-4129 & 13 54 48.47& -41 29 04.8&   0.690& -0.592& -0.354& 14.70& 0.022&   19.4& 18.1& 16.8& 15.52& 15.06& 14.63             & sd\\
PM J14005-3935 & 14 00 32.32& -39 35 29.3&   0.508& -0.496& -0.109& 14.70& 0.022&   17.0& 15.8& 14.1& 13.47& 12.90& 12.66             & sd\\
PM J14104-5001 & 14 10 29.10& -50 01 59.5&   0.538& -0.158& -0.514& 14.19& 0.023&\nodata& 17.1& 14.7& 12.72& 12.31& 11.97          &  d\\
PM J14123-3941 & 14 12 21.16& -39 41 33.4&   0.613& -0.538& -0.293& 16.14& 0.020&   16.3& 14.3& 12.3& 10.99& 10.43& 10.18             &  d\\
PM J14129-4001 & 14 12 58.26& -40 01 21.7&   0.450& -0.414& -0.176& 16.14& 0.020&   20.8& 18.1& 17.8& \nodata& \nodata& \nodata       &  \nodata\\
PM J14330-3846 & 14 33 03.36& -38 46 59.5&   0.475& -0.465& -0.095& 16.14& 0.020&   18.6& 16.0& 15.3& 14.37& 13.78& 13.59             & sd\\
PM J14344-4700 & 14 34 27.25& -47 00 15.6&   0.517& -0.517& -0.002& 18.09& 0.018&   21.1& 15.2& \nodata& 11.94& 11.47& 11.20          &  d\\
PM J14373-4002 & 14 37 21.44& -40 02 49.1&   0.509& -0.408& -0.304& 13.80& 0.023&   15.9& 14.3& 12.4& 10.79& 10.21&  9.90             &  d\\
PM J14382-6231 & 14 38 13.58& -62 31 39.5&   0.695& -0.546& -0.429& 17.34& 0.018&\nodata& 13.4& 12.8& 10.52&  9.98&  9.72          &  d\\
PM J14435-5430 & 14 43 35.13& -54 30 44.6&   0.645& -0.631& -0.131& 15.19& 0.021&   18.0& 15.8& 13.9& 11.96& 11.41& 11.14             &  d\\
PM J14441-3426 & 14 44 06.57& -34 26 46.5&   0.493& -0.030& -0.492& 13.98& 0.023&\nodata& 13.7& 11.0&  9.74&  9.18&  8.88          &  d\\
PM J14440-4414 & 14 44 01.97& -44 14 40.8&   0.509& -0.441& -0.254& 18.09& 0.018&   20.5& 17.0& 15.1& 13.33& 12.81& 12.50             &  d\\
PM J14500-3742 & 14 50 02.87& -37 42 09.6&   0.511& -0.297& -0.416& 13.80& 0.023&   15.0& 13.3& 11.0&  9.95&  9.37&  9.07             &  d\\
PM J14558-3914 & 14 55 51.61& -39 14 33.1&   0.786& -0.780& -0.095& 13.80& 0.023&   20.7& 14.7& \nodata& 12.50& 11.98& 11.79          & sd\\
PM J14570-4705 & 14 57 05.32& -47 05 25.8&   0.548& -0.411& -0.362& 10.34& 0.031&   16.7& 15.2& 14.5& 13.53& 12.96& 12.82             & sd\\
PM J14570-5943 & 14 57 02.97& -59 43 46.8&   0.617& -0.407& -0.463& 17.34& 0.018&   17.9& 15.7& 13.9& 12.48& 12.01& 11.78             &  d\\
PM J14585-5916 & 14 58 31.22& -59 16 42.3&   0.472& -0.423& -0.210& 18.02& 0.018&\nodata& 13.8& \nodata& 10.48&  9.85&  9.58       &  d\\
PM J14596-4043 & 14 59 40.72& -40 43 19.6&   0.754& -0.687& -0.308& 13.87& 0.023&   17.9& 16.6& 14.5& 14.29& 13.80& 13.47             & sd\\
PM J15044-4353 & 15 04 29.26& -43 53 43.8&   0.478& -0.414& -0.240& 10.34& 0.031&   16.8& 15.4& 13.1& 11.79& 11.18& 10.88             &  d\\
PM J15054-4620 & 15 05 27.37& -46 20 16.1&   0.530& -0.462& -0.259& 10.34& 0.031&   16.0& 14.5& \nodata& 11.07& 10.51& 10.28          &  d\\
PM J15116-3403 & 15 11 38.66& -34 03 15.8&   0.567& -0.165& -0.543& 16.25& 0.020&   15.7& 13.7& 11.3& 10.05&  9.42&  9.13             &  d\\
PM J15128-4354 & 15 12 52.34& -43 54 12.4&   0.474& -0.244& -0.407& 10.34& 0.031&   16.2& 13.6& 11.5& 10.57&  9.96&  9.75             &  d\\
PM J15145-4625 & 15 14 31.88& -46 25 55.3&   0.539& -0.538& -0.005& 17.38& 0.018&   15.6& 14.9& 14.4& 14.71& 14.61& 14.65             & wd\\
PM J15184-3544 & 15 18 29.68& -35 44 11.2&   0.481& -0.067& -0.477& 16.25& 0.020&\nodata& 15.8& 14.5& 13.73& 13.28& 13.10          & sd\\
PM J15206-7137 & 15 20 41.70& -71 37 08.3&   0.478& -0.396& -0.267& 16.33& 0.020&   18.0& 16.8& \nodata& 13.25& 12.75& 12.46          &  d\\
PM J15231-7711 & 15 23 11.17& -77 11 23.5&   0.831& -0.772& -0.305& 17.98& 0.018&\nodata& 15.8& 14.7& 13.76& 13.27& 13.03          & sd\\
PM J15246-6239 & 15 24 40.53& -62 39 27.1&   0.497& -0.394& -0.303& 17.06& 0.019&\nodata& 13.9& 12.4& 10.93& 10.48& 10.18          &  d\\
PM J15252-4549 & 15 25 15.12& -45 49 36.5&   0.457& -0.202& -0.410& 17.38& 0.018&   18.4& 16.4& 15.3& 14.49& 14.02& 13.76             & sd\\
PM J15292-5620 & 15 29 15.30& -56 20 40.6&   0.598& -0.576& -0.158& 17.04& 0.019&   17.8& 15.6& 14.5& 13.23& 12.73& 12.48             & sd\\
PM J15322-3622 & 15 32 13.92& -36 22 30.7&   0.462& -0.372& -0.274& 15.85& 0.020&\nodata& 13.0& \nodata& 10.10&  9.54&  9.28       &  d\\
PM J15334-3634 & 15 33 27.72& -36 34 02.4&   0.552& -0.453& -0.316& 15.85& 0.020&   16.4& 14.2& 12.5& 11.54& 10.99& 10.76             &  d\\
PM J15345-6452 & 15 34 31.45& -64 52 54.8&   0.455& -0.139& -0.434& 17.06& 0.019&   15.9& 14.1& 12.4& 11.19& 10.65& 10.39             &  d\\
PM J15413-3609 & 15 41 19.05& -36 09 10.8&   0.529& -0.520& -0.094& 15.85& 0.020&   17.6& 16.1& 13.5& 11.97& 11.45& 11.11             &  d\\
PM J15547-3856 & 15 54 46.12& -38 56 17.4&   0.454& -0.203& -0.406& 17.82& 0.018&   16.2& 13.6& 11.8& 11.19& 10.69& 10.46             &  d\\
PM J16006-3234 & 16 00 39.59& -32 34 11.4&   0.534& -0.443& -0.298& 14.22& 0.023&   20.2& 17.3& 15.9& 14.43& 14.04& 13.69             & sd\\
PM J16019-3421 & 16 01 55.61& -34 21 56.4&   0.672&  0.586& -0.328& 14.22& 0.023&\nodata& 15.0& 12.5& 10.96& 10.33&  9.98          &  d\\
PM J16087-4442 & 16 08 43.95& -44 42 28.3&   0.596& -0.142& -0.579& 17.89& 0.018&   20.2& 14.3& \nodata& 10.88& 10.35& 10.10          &  d\\
PM J16138-3040 & 16 13 53.61& -30 40 57.9&   0.457& -0.304& -0.341& 14.72& 0.022&   16.5& 15.1& 13.5& 13.15& 12.58& 12.38             & sd\\
PM J16141-4044 & 16 14 07.03& -40 44 15.0&   0.523& -0.433& -0.292& 17.82& 0.018&   16.8& 14.5& 13.1& 12.33& 11.87& 11.61             & sd\\
PM J16169-4352 & 16 16 58.36& -43 52 17.1&   0.496& -0.486& -0.099& 16.84& 0.019&   17.5& 15.6& 13.9& 12.28& 11.78& 11.48             &  d\\
PM J16256-6712 & 16 25 38.95& -67 12 21.4&   0.698& -0.561& -0.415& 14.08& 0.023&   18.1& 16.8& 14.3& 12.45& 11.95& 11.60             &  d\\
PM J16364-8737 & 16 36 26.10& -87 37 06.2&   0.451& -0.309& -0.328& 18.67& 0.017&   17.8& 17.2& 17.4& 16.59& 16.26& 16.75             & wd\\
PM J16492-5142 & 16 49 13.81& -51 42 52.2&   0.452& -0.262& -0.369& 16.31& 0.020&   15.4& 14.4& 11.9& 12.12& 11.63& 11.44             & sd\\
PM J17037-6823 & 17 03 46.81& -68 23 29.3&   0.503& -0.347& -0.363& 15.02& 0.021&   16.8& 14.6& 13.1& 13.13& 12.65& 12.47             & sd\\
PM J17056-8608 & 17 05 38.35& -86 08 44.1&   0.512& -0.305& -0.411& 18.67& 0.017&   19.2& 16.9& 13.6& 12.82& 12.28& 11.96             &  d\\
PM J17397-6322 & 17 39 46.10& -63 22 07.1&   0.547& -0.098& -0.538& 17.07& 0.019&\nodata& 13.3& 11.8& 11.08& 10.50& 10.25          &  d\\
PM J18004-6358 & 18 00 29.82& -63 58 21.4&   0.472& -0.183& -0.435& 17.07& 0.019&   18.4& 16.1& 14.8& 14.40& 13.94& 13.72             & sd\\
PM J18036-6340 & 18 03 37.17& -63 40 47.9&   0.484& -0.256& -0.411& 17.27& 0.019&   17.2& 15.5& 14.3& 13.67& 13.13& 12.92             & sd\\
PM J18217-6101 & 18 21 46.25& -61 01 54.5&   0.559&  0.074& -0.554& 19.99& 0.016&   20.3& 18.1& 14.6& 12.82& 12.29& 11.96             &  d\\
PM J18286-4531 & 18 28 38.23& -45 31 08.1&   0.501& -0.331& -0.376& 15.92& 0.020&   19.0& 14.7& 15.3& 12.24& 11.72& 11.43             & sd\\
PM J18298-5131 & 18 29 51.30& -51 31 31.7&   0.577& -0.241& -0.524& 15.04& 0.021&   19.5& 18.0& \nodata& 15.63& 14.95& 14.79          & sd\\
PM J19105-4132 & 19 10 33.60& -41 32 50.6&   0.676&  0.125& -0.664& 14.95& 0.021&\nodata& \nodata& \nodata& 11.14& 10.55& 10.24    &  d\\
PM J19105-4133 & 19 10 34.60& -41 33 44.4&   0.676&  0.125& -0.664& 14.95& 0.021&\nodata& 13.0& 11.0&  9.85&  9.24&  9.03          &  d\\
PM J19103-4338 & 19 10 23.58& -43 38 37.2&   0.450& -0.004& -0.450& 14.74& 0.022&\nodata& 15.7& 13.7& 11.86& 11.28& 10.99          &  d\\
PM J19110-3820 & 19 11 00.24& -38 20 31.9&   0.585&  0.547& -0.205& 14.95& 0.021&\nodata& 16.9& \nodata& 15.99& 15.79& 15.62       & wd\\
PM J19167-3638 & 19 16 46.58& -36 38 04.1&   1.340& -0.162& -1.329& 15.04& 0.021&   18.1& 15.8& 14.9& 13.66& 13.12& 12.95             & sd\\
PM J19184-4554 & 19 18 29.44& -45 54 30.8&   0.616& -0.414& -0.456& 14.74& 0.022&\nodata& 15.1& 13.1& 11.21& 10.65& 10.30          &  d\\
PM J19248-3356 & 19 24 48.28& -33 56 09.7&   0.574&  0.319& -0.477& 15.04& 0.021&   16.0& 13.7& 12.8& 12.45& 11.99& 11.77             & sd\\
PM J19261-4310 & 19 26 08.60& -43 10 56.4&   1.142& -0.142& -1.133& 16.96& 0.019&\nodata& 15.8& 13.7& 11.94& 11.42& 11.12          &  d\\
PM J19285-3634 & 19 28 33.61& -36 34 30.3&   0.503&  0.024& -0.503& 14.72& 0.022&\nodata& 14.2& 12.4& 10.61& 10.06&  9.81          &  d\\
PM J19280-6028 & 19 28 00.48& -60 28 52.4&   0.460&  0.068& -0.455& 13.77& 0.023&\nodata& 16.6& 14.5& 13.97& 13.41& 13.23          & sd\\
PM J19403-3944 & 19 40 21.30& -39 44 10.0&   0.529&  0.156& -0.505& 14.20& 0.023&\nodata& 13.8& 17.4& 10.38&  9.84&  9.57          &  d\\
PM J20444-4123 & 20 44 27.89& -41 23 51.6&   0.505&  0.357& -0.356& 15.25& 0.021&\nodata& 14.3& 12.4& 11.75& 11.16& 10.99          &  d\\
PM J20460-5458 & 20 46 01.48& -54 58 06.7&   0.461&  0.134& -0.441& 15.04& 0.021&\nodata& 18.5& 16.7& 15.29& 14.80& 14.41          & sd\\
PM J20530-5409 & 20 53 03.90& -54 09 37.5&   0.633&  0.177& -0.608& 15.04& 0.021&   18.7& 16.6& 13.8& 12.17& 11.66& 11.35             &  d\\
PM J21324-3922 & 21 32 29.67& -39 22 50.1&   0.513&  0.447& -0.252& 16.08& 0.020&\nodata& 15.8& 13.4& 12.21& 11.70& 11.35          &  d\\
PM J22040-3347 & 22 04 02.27& -33 47 38.3&   0.948&  0.472& -0.822& 14.14& 0.023&\nodata& 14.5& \nodata& 12.32& 11.81& 11.60       & sd\\
PM J22178-7753 & 22 17 53.89& -77 53 36.0&   0.584&  0.315& -0.492& 15.92& 0.020&   18.9& 17.0& 15.5& 14.53& 14.07& 13.91             & sd\\
PM J22209-3346 & 22 20 57.91& -33 46 57.9&   0.559& -0.102& -0.550& 15.12& 0.021&   19.8& 17.8& 18.1& \nodata& \nodata& \nodata       &  \nodata\\
PM J22359-7722 & 22 35 57.74& -77 22 16.8&   0.601& -0.224& -0.558& 15.92& 0.020&   18.4& 16.3& 15.3& 14.17& 13.67& 13.46             & sd\\
PM J22387-6232 & 22 38 47.28& -62 32 22.7&   0.504&  0.305& -0.401& 18.11& 0.018&   20.3& 17.4& 18.2& \nodata& \nodata&  \nodata      &  \nodata\\
PM J22403-4931E& 22 40 18.96& -49 31 01.4&   0.504&  0.457&  0.212& 14.87& 0.022&   15.1& 13.3& 10.7&  9.93&  9.38&  9.03             &  d\\
PM J22403-4931W& 22 40 18.67& -49 31 04.6&   0.504&  0.457&  0.212& 14.87& 0.022&\nodata& \nodata& \nodata&  9.84&  9.26&  9.01    &  d\\
PM J23350-4904 & 23 35 04.31& -49 04 54.3&   0.611&  0.334& -0.511& 17.33& 0.018&\nodata& 15.5& 13.9& 13.29& 12.75& 12.52	        sd
\enddata                              
\tablenotetext{a}{Temporal baseline between the first and second DSS epoch.}
\tablenotetext{b}{Estimated proper motion error.}
\tablenotetext{c}{Photographic $B_J$ (IIIaJ), $R_F$ (IIIaF) and $I_N$
(IVN) magnitudes from the USNO-B1.0 catalog.}
\tablenotetext{d}{Infrared $JHK_s$ magnitudes from the
2MASS All-Sky Point Source Catalog.}
\end{deluxetable}                     
\clearpage
\end{landscape}

\end{document}